\documentclass[aps,prl,twocolumn,reprint,superscriptaddress,citeautoscript]{revtex4-2}

\usepackage{graphicx}
\usepackage{dcolumn}
\usepackage{bm}

\usepackage{xcolor}
\usepackage{soul}
\usepackage{graphicx}
\usepackage{dcolumn}
\usepackage{bm}
\usepackage{mathrsfs}
\usepackage{amsmath}
\def\lesssim{\ \raise.3ex\hbox{$<$}\kern-0.8em\lower.7ex\hbox{$\sim$}\ }
\def\gesim{\ \raise.3ex\hbox{$>$}\kern-0.8em\lower.7ex\hbox{$\sim$}\ }

\begin{document}
\title{Tunable Fujita-Miyazawa-Type Three-Body Force in Ultracold Atoms}
\author{Hiroyuki Tajima}
\affiliation{Department of Physics, Graduate School of Science, The University of Tokyo, Tokyo 113-0033, Japan}
\affiliation{RIKEN Nishina Center, Wako 351-0198, Japan}
\author{Eiji Nakano}
\affiliation{Department of Mathematics and Physics, Kochi University, Kochi 780-8520, Japan}
\author{Kei Iida}
\affiliation{Department of Liberal Arts, The Open University of Japan, Chiba 261-8586, Japan}
\affiliation{Department of Mathematics and Physics, Kochi University, Kochi 780-8520, Japan}
\affiliation{RIKEN Nishina Center, Wako 351-0198, Japan}

\date{\today}
\begin{abstract}
  We show how a Fujita-Miyazawa-type three-body force emerges among three impurity atoms immersed in an atomic Bose-Einstein condensate near an interspecies Feshbach resonance. 
  As a result of thermal average over excitations in the medium and impurities as well as expansion with respect to the impurity-medium and Feshbach resonance couplings, two superfluid phonons and a closed channel resonance play a role in producing an effective three-body force, as in the original three-nucleon case in which two pions and a $\Delta$ resonance are involved. The proposed Fujita-Miyazawa-type three-body force can be enhanced by tuning the closed-channel energy level via an external magnetic field, and moreover, its strength can be confirmed experimentally by measuring the impurity equation of state.
Our result gives a new insight into an analogy between atomic polarons and nuclear few-body systems.
\end{abstract}
\maketitle

{\it Introduction.---}
Three or more-body interactions necessarily appear among compound particles, but are typically too weak to be qualitatively relevant because the excitation energies of each compound particle are extremely higher than the typical kinetic energy.
As long as we are interested in many-particle phenomena of which the energy scale is of the order of the kinetic energy,
we basically have only to take into account two-body interactions, which often give rise to phase transitions such as
superconductivity, crystallization, and condensation.

If two-body repulsion and attraction counteract each other, however, a three-body force could play a significant role.  Indeed, the saturation of the density and binding energy of atomic nuclei is predicted to require a significant correction due to the three-nucleon force~\cite{PhysRevC.58.1804,machleidt2019wrong}.
This is natural partly because a triton has a significant fraction of the binding energy contributed by the three-nucleon force
according to existing model-dependent few-body analyses \cite{kalantar2011signatures} and partly because
nonnegligible three-body-interaction effects on the nucleon-deuteron elastic scattering have been empirically revealed~\cite{PhysRevLett.81.1183,PhysRevC.65.034003,PhysRevC.70.014001,PhysRevLett.95.162301,PhysRevC.76.014004,PhysRevC.89.064007,PhysRevC.96.064001}.

\begin{figure}[t]
    \centering
    \includegraphics[width=1.0\linewidth]{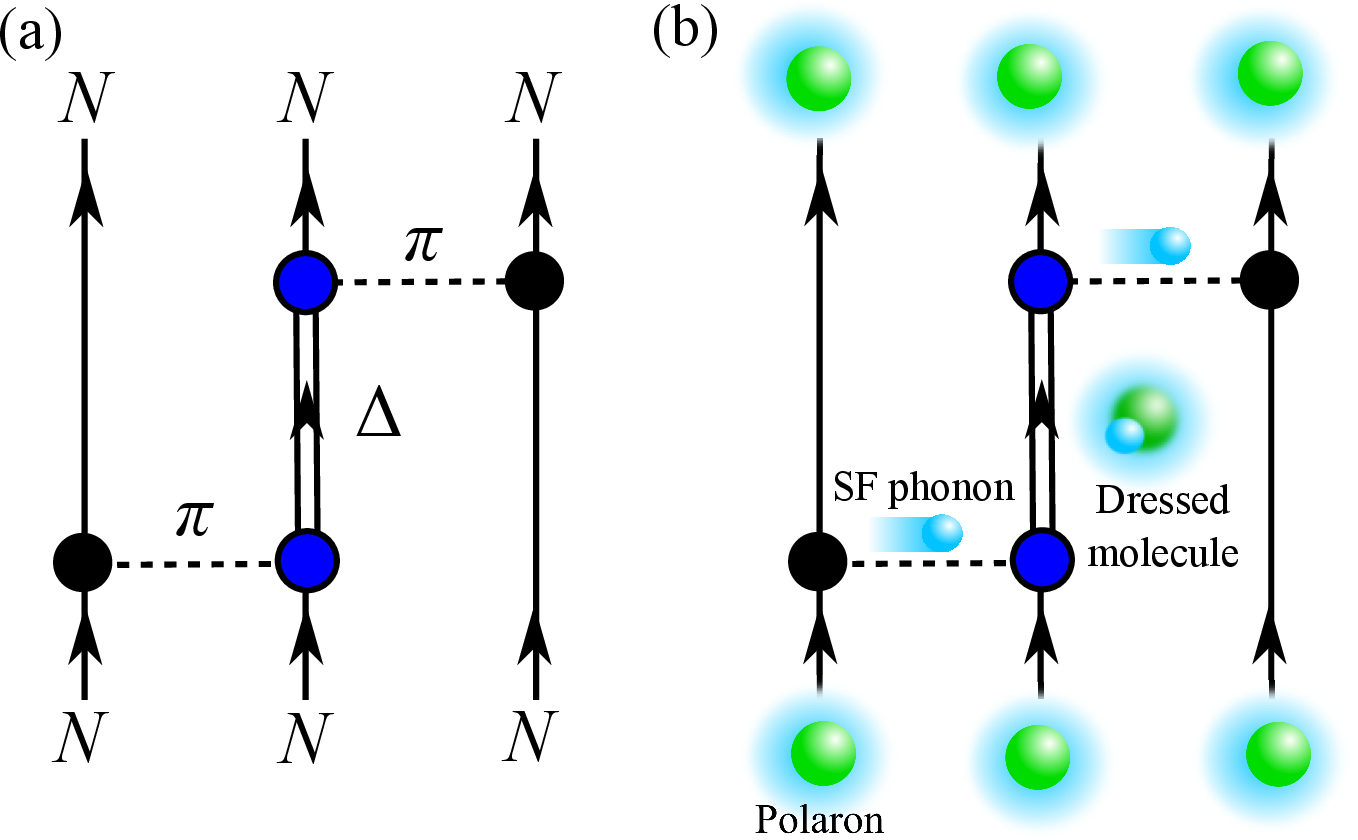}
    \caption{(a) Feynman diagram representing the Fujita-Miyazawa three-body force, which acts on three dynamical nucleons ($N$)
       via a two-pion ($\pi$) exchange process involving a virtual excited state ($\Delta$). 
       (b) Schematics for the analog three-body force in an ultracold atomic mixture near the Feshbach resonance. The Fujita-Miyazawa-type three-body force emerges
       among three $N$-like polarons in the condensate, which is accompanied by exchange of two $\pi$-like superfluid (SF) phonons
       and a virtual excitation of a $\Delta$-like dressed closed-channel molecule.
       }
    \label{fig:1}
\end{figure}

In particular, a three-nucleon force proposed by
Fujita and Miyazawa laid the foundation for later developments in the study of multi-body nuclear forces~\cite{10.1143/PTP.17.360}.
As shown in Fig.~\ref{fig:1}(a),
the Fujita-Miyazawa three-body force is induced by a two-pion exchange process involving a virtual $\Delta$ state.
Since experimental data that constrain the Fujita-Miyazawa three-nucleon interaction exclusively are still limited,
it is interesting to study the analog of the Fujita-Miyazawa three-body force in other controllable systems.

To this end, an ultracold atomic gas is promising,
since its physical parameters can be changed ~\cite{bloch2012quantum}
in such a way as to tune the two-body interaction via magnetic Feshbach resonances~\cite{RevModPhys.82.1225}.
Indeed, the three-body interaction has been explored extensively in this atomic system.
The idea that an effective three-body interaction should be induced by virtual excitations to low-lying vibrational states in a lattice potential has been proposed~\cite{johnson2009effective,johnson2012effective} and demonstrated in recent experiments~\cite{goban2018emergence,PhysRevA.111.033303}.
Also, the Rabi coupling in binary mixtures induces a three-body interaction among ground-state dressed states
~\cite{PhysRevA.90.021601,PhysRevA.97.011602,tiengo2025three}, where effects of excited dressed states are built into the multi-body interactions.
This scheme has been realized experimentally~\cite{PhysRevLett.128.083401}.

Another example of the three-body interaction in ultracold atoms is the effective one among polarons~\cite{scazza2022repulsive,baroni2024quantum,tajima2024intersections}.
Impurity atoms immersed in a medium gas are dressed with medium excitations and exchange them with each other, resulting in mediated interactions~\cite{PhysRevA.110.030101}.
Remarkably, exchange of a single superfluid phonon induces a Yukawa-type two-body interaction~\cite{yukawa1935interaction} between impurities immersed in a Bose-Einstein condensate (BEC)~\cite{PhysRevLett.85.2418,PhysRevA.61.053601,
PhysRevB.93.205144,naidon2018two,PhysRevLett.121.013401},
indicating a close analogy between interpolaron interactions and the nuclear force.
This analogy looks very convincing because each nucleon is not a bare particle but accompanied by a pionic cloud just like a polaron~\cite{miyazima1942sc,TomonagaPTP.1.83,TomonagaPTP.1.109,tomonaga1947effect,TomonagaPTP.2.63}.
Indeed, the effective three-body interaction induced by 
medium-fermion exchange~\cite{PhysRevA.69.043607,PhysRevA.76.013609,friman2011three,PhysRevA.104.L041302,tajima2021polaron,PhysRevC.104.065801,PhysRevA.109.013319} has been studied theoretically, while its two-body counterpart~\cite{desalvo2019observation,PhysRevLett.124.163401,baroni2024mediated,cai2025fermion}
and also a residual three-body interaction~\cite{PhysRevLett.131.083003}
have been observed experimentally.
However, it has not been clarified whether the analog of the Fujita-Miyazawa three-body force among polarons emerges or not.

In this work, we consider a Fujita-Miyazawa-type three-body force among three impurities immersed in a BEC near the intercomponent Feshbach resonance.
This system, if described by a two-channel model relevant to the narrow Feshbach resonance, is advantageous for mimicking both nucleons and $\Delta$ resonance in terms of the superposition of open- and closed-channel states.
It should be noted that the seminal paper~\cite{10.1143/PTP.17.360} by Fujita and Miyazawa is the first to consider three-body interactions involving $\Delta$ resonance
as an explicit isobaric degree of freedom, which is now systematically incorporated in the coupled-channel treatment (for instance, see Refs.~\cite{PhysRevC.48.2190,PhysRevC.68.024005,machleidt2011chiral}). 
Incidentally, the mediated interaction in the two-channel model has been found to involve non-trivial features such as pair-exchange coupling~\cite{PhysRevA.108.023304}, which cannot be described in a single-channel model.
As depicted in Fig~\ref{fig:1}(b),
a Fujita-Miyazawa-type three-body force emerges through one superfluid phonon exchange, a subsequent virtual excitation of a dressed closed-channel
molecule, and finally another superfluid phonon exchange.  As for a theoretical framework that will be adopted here,
 we take a trace of the full density matrix over excited states to obtain an effective system of dressed impurities with mediated multi-body interactions.
 This approach has the advantage that one can directly use a perturbation theory within the standard operator formalism without resorting to the path integral formalism.

{\it Model.---}
In what follows, we take $\hbar=k_{\rm B}=1$ and the system volume is set to unity for simplicity.
The two-channel Hamiltonian of a binary quantum mixture near the Feshbach resonance reads
\begin{align}
    &\hat{H}=\sum_{\bm{k}}
    \left[\xi_{\bm{k},b}\hat{b}_{\bm{k}}^\dag \hat{b}_{\bm{k}}
    +\xi_{\bm{k},c}\hat{c}_{\bm{k}}^\dag \hat{c}_{\bm{k}}
    \right]
    +\sum_{\bm{P}}\xi_{\bm{P},A}\hat{A}_{\bm{P}}^\dag \hat{A}_{\bm{P}}\cr
    &+\frac{ U_{bb}}{2}
    \sum_{\bm{k},\bm{k}',\bm{P}}
    \hat{b}_{\bm{k}+\frac{\bm{P}}{2}}^\dag
    \hat{b}_{-\bm{k}+\frac{\bm{P}}{2}}^\dag
    \hat{b}_{-\bm{k}'+\frac{\bm{P}}{2}}
    \hat{b}_{\bm{k}'+\frac{\bm{P}}{2}}\cr
    &+
    U_{bc}
    \sum_{\bm{k},\bm{k}',\bm{P}}
    \hat{b}_{\bm{k}+\frac{M_b}{M_A}\bm{P}}^\dag
    \hat{c}_{-\bm{k}+\frac{M_c}{M_A}\bm{P}}^\dag
    \hat{c}_{-\bm{k}'+\frac{M_c}{M_A}\bm{P}}
    \hat{b}_{\bm{k}'+\frac{M_b}{M_A}\bm{P}}\cr
    &+g\sum_{\bm{P},\bm{k}}
    \left[\hat{A}_{\bm{P}}^\dag \hat{b}_{-\bm{k}+\frac{M_b}{M_A}\bm{P}}\hat{c}_{\bm{k}+\frac{M_c}{M_A}\bm{P}}+{\rm h.c.}\right],
\end{align}
where 
$\xi_{\bm{k},b}=k^2/(2M_b)-\mu_b$,
$\xi_{\bm{k},c}=k^2/(2M_c)-\mu_c$,
and $\xi_{\bm{P},A}=P^2/(2M_A)-\mu_b-\mu_c+\nu$ are the kinetic energies of a medium atom $b$ with mass $M_b$ and chemical potential $\mu_b$, 
an impurity atom $c$ with mass $M_c$ and chemical potential $\mu_c$, and a closed-channel molecule $A$ with mass $M_A=M_c+M_b$ and energy level $\nu$, respectively.
$\hat{b}_{\bm{k}}$, $\hat{c}_{\bm{k}}$, and $\hat{A}_{\bm{P}}$ are the corresponding annihilation operators.
While we do not specify the statistics of the impurity atom and the closed-channel molecule,
the resulting three-body interaction does not depend on whether impurities are fermions or bosons.
$U_{bb}$ and $U_{bc}$ are the background $b$--$b$ and $b$--$c$ interactions,
and $g$ is the Feshbach atom-molecule ($bc$--$A$) coupling.
The direct impurity-impurity ($c$--$c$) interaction is not shown here because it is not important for our purpose.

\begin{figure}[t]
    \centering
    \includegraphics[width=1\linewidth]{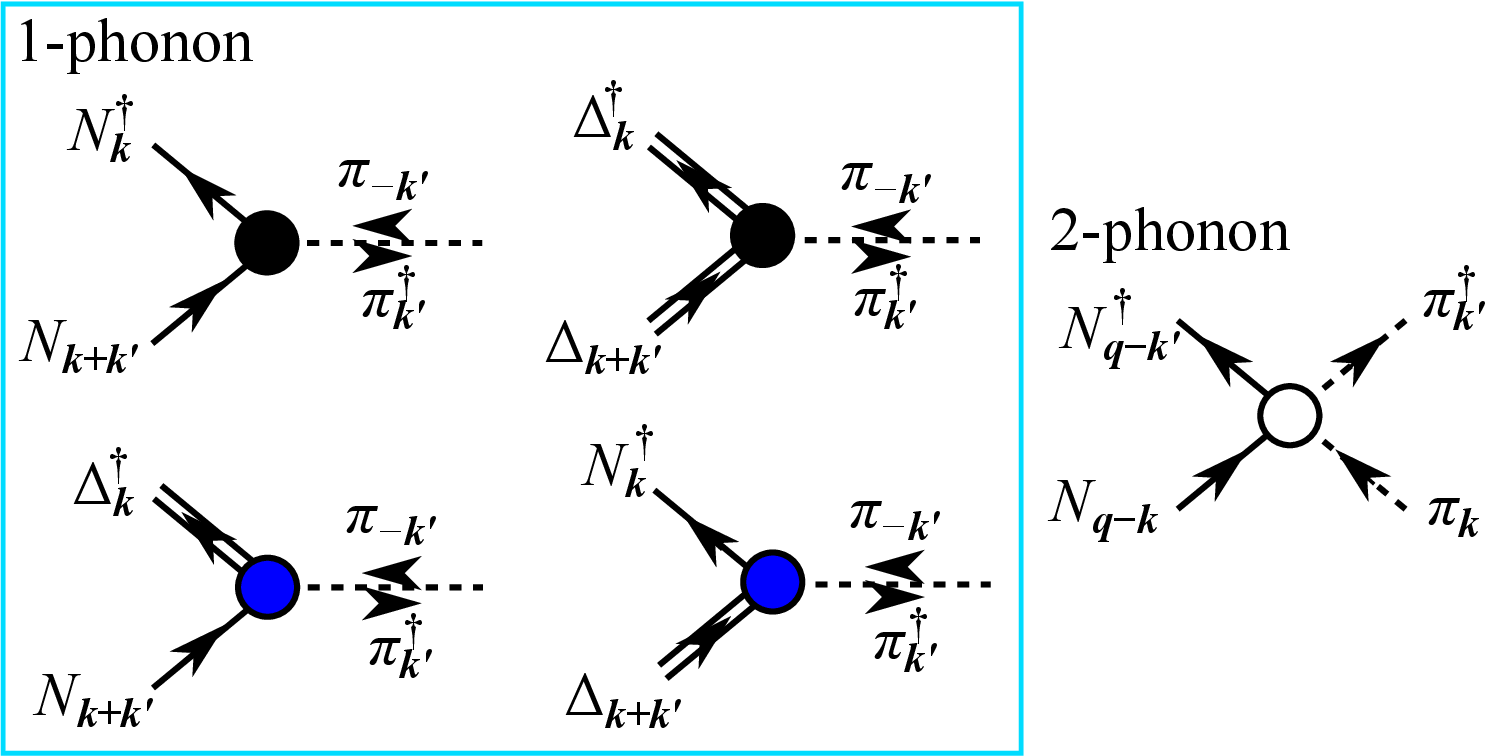}
    \caption{Diagrammatic representation of the interactions involving the one superfluid phonon process given by Eq.~\eqref{eq:5}. For reference, we also show an example of the two-phonon process. The filled and open circles represent the one- and two-phonon vertices, respectively.}
    \label{fig:2}
\end{figure}

Let us now separate $\hat{b}_{\bm{k}}$ into the condensation and excitation parts
as
$\hat{b}_{\bm{k}}=\sqrt{n_0}\delta_{\bm{k},\bm{0}}+\hat{\pi}_{\bm{k}}(1-\delta_{\bm{k},\bm{0}})$,
where $n_0$ is the condensate density.
We then apply the Bogoliubov approximation for medium excitations and for impurity and molecule states,
which allows us to rewrite the Hamiltonian as
$\hat{H}=
\hat{H}_N+\hat{H}_\Delta
+\hat{H}_{\pi}+\hat{V}$,
where
\begin{align}
    &\hat{H}_{N}=\sum_{\bm{k}}
    \xi_{\bm{k},N}
    \hat{N}_{\bm{k}}^\dag
    \hat{N}_{\bm{k}}
    ,
    &\hat{H}_{\Delta}=\sum_{\bm{k}}
    \xi_{\bm{k},\Delta}
    \hat{\Delta}_{\bm{k}}^\dag
    \hat{\Delta}_{\bm{k}}
    ,
\end{align}
are the kinetic terms of dressed nucleon ($N$)-like impurity states $\hat{N}_{\bm{k}}=-s_{\bm{k}}^{-}\hat{c}_{\bm{k}}+s_{\bm{k}}^{+}\hat{A}_{\bm{k}}$ and $\Delta$-like excited states $\hat{\Delta}_{\bm{k}}=s_{\bm{k}}^{+}\hat{c}_{\bm{k}}+s_{\bm{k}}^{-}\hat{A}_{\bm{k}}$
induced by the Rabi-type mixing term $g\sqrt{n}_0\sum_{\bm{k}}\left(c_{\bm{k}}^\dag A_{\bm{k}}
+A_{\bm{k}}^\dag c_{\bm{k}}
\right)$ as polarons with weights
\begin{align}
    s_{\bm{k}}^{\pm}=\sqrt{\frac{1}{2}\left(1\pm\frac{\xi_{\bm{k},{c}}^*-\xi_{\bm{k},{A}}}{\sqrt{(\xi_{\bm{k},{c}}^*-\xi_{\bm{k},{ A}})^2+4g^2n_0}}\right)}.
\end{align}
Here, 
the kinetic energies of the diagonalized $N$-like and $\Delta$-like polaronic states are given by
    $\xi_{\bm{k},N}=
    \left(\xi_{\bm{k},{c}}^*+\xi_{\bm{k},{A}}
    -\sqrt{(\xi_{\bm{k},{c}}^*-\xi_{\bm{k},{A}})^2+4g^2n_0}
    \right)/
    {2}$ and
    $\xi_{\bm{k},\Delta}=
    \left(\xi_{\bm{k},{c}}^*+\xi_{\bm{k},{A}}
    +\sqrt{(\xi_{\bm{k},{c}}^*-\xi_{\bm{k},{A}})^2+4g^2n_0}
    \right)
    /{2}$,
where the impurity kinetic energy involves the Hartree shift as $\xi_{\bm{k},c}^*=\xi_{\bm{k},c}+U_{bc}n_0$.    
    Note that $\xi_{\bm{k},\Delta}\geq \xi_{\bm{k},N}$ for any value of $\nu$.
    $\hat{H}_\pi$ is the conventional Bogoliubov Hamiltonian given by
\begin{align}
    \hat{H}_\pi
    &=\sum_{\bm{k}}(\xi_{\bm{k},b}+2U_{bb}n_0)\hat{\pi}_{\bm{k}}^\dag  \hat{\pi}_{\bm{k}}\cr
    &
    +\frac{U_{bb}n_0}{2}
    \sum_{\bm{k}}\left[\hat{\pi}_{\bm{k}}^\dag \hat{\pi}_{-\bm{k}}^\dag +\hat{\pi}_{-\bm{k}}\hat{\pi}_{\bm{k}}\right] + {\rm const.},
\end{align}
     which, once diagonalized, leads to the kinetic term of superfluid phonons,
      i.e., superposed states of the particle and hole associated with $\hat{\pi}_{\bm{k}}^\dag$ and $\hat{\pi}_{-\bm{k}}$.
Finally, $\hat{V}$ represents the interaction term in the form of
\begin{align}
\label{eq:5}
        &\hat{V}=
    \sum_{\bm{k},\bm{k}'}\left[
    f_{\bm{k},\bm{k}'}^{NN\pi}
    \hat{N}_{\bm{k}}^\dag
    \hat{N}_{\bm{k}+\bm{k}'}\hat{\pi}_{\bm{k}'}^\dag 
    +
    f_{\bm{k},\bm{k}'}^{\Delta\Delta\pi}
    \hat{\Delta}_{\bm{k}}^\dag
    \hat{\Delta}_{\bm{k}+\bm{k}'}\hat{\pi}_{\bm{k}'}^\dag\right.  \cr
    &
    \,\,\,
    \left.
    +
    f_{\bm{k},\bm{k}'}^{\Delta N\pi}
    \hat{\Delta}_{\bm{k}}^\dag
    \hat{N}_{\bm{k}+\bm{k}'}\hat{\pi}_{\bm{k}'}^\dag  
    +
        f_{\bm{k},\bm{k}'}^{N\Delta \pi}
    \hat{N}_{\bm{k}}^\dag
    \hat{\Delta}_{\bm{k}+\bm{k}'}
    \hat{\pi}_{\bm{k}'}^\dag\right]
    +{\rm h.c.},
\end{align}
which describes the absorption and emission of one superfluid phonon as shown in Fig.~\ref{fig:2}.
The form factors at the one-phonon vertices are given by
\begin{align}
f_{\bm{k},\bm{k}'}^{NN\pi}&=s^{-}_{\bm{k}}s^{-}_{\bm{k}+\bm{k}'}U_{bc}\sqrt{n_0}
    -s^{-}_{\bm{k}}s^{+}_{\bm{k}+\bm{k}'}g,\cr
    f_{\bm{k},\bm{k}'}^{\Delta\Delta\pi}&=
    s^{+}_{\bm{k}}s^{+}_{\bm{k}+\bm{k}'}U_{bc}\sqrt{n_0}
    +s^{+}_{\bm{k}}s^{-}_{\bm{k}+\bm{k}'}g,\cr
    f_{\bm{k},\bm{k}'}^{\Delta N\pi}&=-s^{+}_{\bm{k}}s^{-}_{\bm{k}+\bm{k}'}U_{bc}\sqrt{n_0}
    +s^{+}_{\bm{k}}s^{+}_{\bm{k}+\bm{k}'}g,\cr
        f_{\bm{k},\bm{k}'}^{N\Delta \pi}&=-s^{-}_{\bm{k}}s^{+}_{\bm{k}+\bm{k}'}U_{bc}\sqrt{n_0}
    -s^{-}_{\bm{k}}s^{-}_{\bm{k}+\bm{k}'}g.
\end{align}
Here we keep the leading-order terms, which do not contain the two-phonon vertex (open circle in Fig.~\ref{fig:2}).
This approach can be justified when the coupling constants $g$ and $U_{bc}$ are sufficiently small for the depletion of the condensate to be negligible~\cite{PhysRevX.8.031042}.
In this case, the total boson density is given by $n\simeq n_0\gg 1$
for the unit volume.
In fact, the vertex for the two-phonon process, which is $O(U_{bc})$,
is negligible compared to the one-phonon case of $O(U_{bc}\sqrt{n_0})$.
The one-phonon process
leads to the celebrated Yukawa-type two-body interaction~\cite{PhysRevLett.85.2418,PhysRevA.61.053601,
PhysRevB.93.205144,naidon2018two,PhysRevLett.121.013401},
whereas the two-phonon process would be important for the van der Waals-type long-range part of the mediated interactions~\cite{PhysRevC.94.055806,PhysRevLett.129.233401}.

{\it Partial-trace approach to the effective interaction.---}
We proceed to a formalism for deriving the effective interaction among $N$-like polarons in the ground state
by tracing out the degrees of freedom of superfluid phonons and $\Delta$-like polarons.
The grand partition function $Z$ of the total system reads
\begin{align}
Z&={\rm Tr}\left[e^{-\beta(\hat{H}_{N}+\hat{H}_{\Delta}+\hat{H}_{\pi}+\hat{V})}\right]\cr
&\equiv {\rm Tr}\left[e^{-\beta(\hat{H}_{N}+\hat{H}_{\Delta}+\hat{H}_{\pi})}\hat{S}(\beta)\right],
\end{align}
where 
\begin{align}
\label{eq:7}
    \hat{S}(\beta)=T_\tau\exp\left[-\int_0^{\beta}d\tau\,\hat{V}(\tau)\right]
\end{align}
is the $S$-matrix operator for $\hat{V}(\tau)=e^{\tau (\hat{H}_N+\hat{H}_\Delta+\hat{H}_\pi)}\hat{V}e^{-\tau (\hat{H}_N+\hat{H}_\Delta+\hat{H}_\pi)}$ with the imaginary time $\tau$.
$T_\tau$ in Eq.~\eqref{eq:7} is the imaginary-time-ordered product.
We are interested in the phonon-mediated effective interaction $\hat{V}_{\rm eff}$,
which can be defined through
\begin{align}
    Z={\rm Tr}_{N}\left[e^{-\beta (\hat{H}_{N}+\hat{V}_{\rm eff})}\right],
\end{align}
where ${\rm Tr}_{N}\left[\cdots\right]$ denotes the partial trace over the $N$ eigenstates.
Then, $\hat{V}_{\rm eff}$ satisfies
\begin{align}
\label{eq:9}
    e^{-\beta \hat{V}_{\rm eff}}
    ={\rm Tr}_{\Delta\pi} \left[e^{-\beta (\hat{H}_\Delta+\hat{H}_\pi)}\hat{S}(\beta)\right],
\end{align}
with the partial trace ${\rm Tr}_{\Delta \pi}\left[\cdots\right]$ with respect to $\Delta$ and $\pi$.  
Expanding $\hat{S}(\beta)$ in terms of $\hat{V}$,
one may obtain the perturbative expression for $\hat{V}_{\rm eff}$ as
\begin{align}
\label{eq:8}
    \hat{V}_{\rm eff}
    &=\sum_{\ell=1}^{\infty}\frac{(-1)^{\ell-1}}{\ell!\beta}
    \int_0^{\beta}d\tau_1\cdots \int_0^{\beta} d\tau_\ell\cr
    &
    \quad\quad\quad\quad
    \times
    \langle T_\tau[\hat{V}(\tau_1)\cdots\hat{V}(\tau_\ell)]\rangle,
\end{align}
where $\langle\cdots\rangle ={\rm Tr}_{\pi\Delta}[e^{-\beta(\hat{H}_\Delta+\hat{H}_\pi)}\cdots]/{\rm Tr}_{\pi\Delta}[e^{-\beta(\hat{H}_\Delta+\hat{H}_\pi)}]$ is the thermal average over the $\pi$ and $\Delta$ eigenstates only for the connected diagrams.
We note that
$\hat{V}_{\rm eff}$ is still an operator acting on the Hilbert space composed of the $N$ states. 
The Bloch--De Dominicis theorem allows us to decompose the thermal average into the Green's functions of superfluid phonons
$G_{11}(\bm{q},\tau)=-\langle T_\tau[\hat{\pi}_{\bm{q}}(\tau)\hat{\pi}_{\bm{q}}^\dag(0)]\rangle$
and $G_{12}(\bm{q},\tau)=-\langle T_\tau[\hat{\pi}_{\bm{q}}^\dag(\tau)\hat{\pi}_{-\bm{q}}^\dag(0)]\rangle$~\cite{RevModPhys.76.599} and the excited $\Delta$-like polaron state $G_\Delta(\bm{k},\tau)=-\langle T_\tau[\hat{\Delta}_{\bm{k}}(\tau)\hat{\Delta}_{\bm{k}}^\dag(0)]\rangle$.

{\it Tunable Fujita-Miyazawa-type three-body force.---}
We can now give an expression for the Fujita-Miyazawa-type three-body force $\hat{V}_{\rm FM}$ among three low-energy $N$-like polarons diagrammatically
shown in Fig.~\ref{fig:1}(b).
For the sake of obtaining the static three-body potential,
we assume that $N$-like polarons remain at rest during the time scales of the dynamics of $\Delta$-like polarons and superfluid phonons. 
This assumption allows us to take the low-frequency limit of the energy transfer between
the incoming and outgoing $N$ states. Then, $\hat{V}_{\rm FM}$ manifests itself as the fourth-order term with respect to $\hat{V}$ in Eq.~\eqref{eq:8}, which reads
\begin{align}
    \hat{V}_{\rm FM}
    &=\frac{1}{6}
    \sum_{\bm{k}_1,\bm{k}_2,\bm{k}_3,\bm{q}_1,\bm{q}_2}
    U_{\bm{k}_1,\bm{k}_2,\bm{k}_3}(\bm{q}_1,\bm{q}_2)\cr
     &\times
        \hat{N}_{\bm{k}_1}^\dag
    \hat{N}_{\bm{k}_2}^\dag
    \hat{N}_{\bm{k}_3}^\dag
    \hat{N}_{\bm{k}_3-\bm{q}_1}
        \hat{N}_{\bm{k}_2+\bm{q}_1-\bm{q}_2}
    \hat{N}_{\bm{k}_1+\bm{q}_2},    
\end{align}
where the coupling strength is given by
\begin{align}
\label{eq:12}
        U_{\bm{k}_1,\bm{k}_2,\bm{k}_3}(\bm{q}_1,\bm{q}_2)
    =
    -6
    \mathcal{G}_{\bm{k}_2+\bm{q}_1}^{\Delta}
    \mathcal{G}_{\bm{k}_1,\bm{q}_2,\bm{k}_2+\bm{q}_1}^{N\pi\Delta}
       \mathcal{G}_{\bm{k}_2,\bm{q}_1,\bm{k}_3}^{\Delta\pi N},
\end{align}
where $\mathcal{G}_{\bm{k}_2+\bm{q}_1}^{\Delta}=-1/\xi_{\bm{k}_2+\bm{q}_1,\Delta}$ is the zero-frequency propagator of the $\Delta$-like impurity. 
$\mathcal{G}_{\bm{k}_1,\bm{q}_2,\bm{k}_2+\bm{q}_1}^{N\pi\Delta}$ and $\mathcal{G}_{\bm{k}_2,\bm{q}_1,\bm{k}_3}^{\Delta\pi N}$ are the superfluid phonon propagators with the form factors given by
\begin{align}
    \mathcal{G}_{\bm{k}_1,\bm{q}_2,\bm{k}_2+\bm{q}_1}^{N\pi\Delta} 
    &=\mathcal{G}_{\bm{q}_2}^{11}
    \left[
     f_{\bm{k}_1+\bm{q}_2,-\bm{q}_2}^{NN\pi}
f_{\bm{k}_2+\bm{q}_1,-\bm{q}_2}^{\Delta N\pi}\right.\cr
&\quad\quad\quad\quad\left.+
 f_{\bm{k}_1,\bm{q}_2}^{NN\pi}
   f_{\bm{k}_2+\bm{q}_1-\bm{q}_2,\bm{q}_2}^{N\Delta\pi}
    \right]\cr
    &+\mathcal{G}_{\bm{q}_2}^{12}
    \left[
     f_{\bm{k}_1,\bm{q}_2}^{NN\pi}
   f_{\bm{k}_2+\bm{q}_1-\bm{q}_2,\bm{q}_2}^{\Delta N\pi}\right.\cr
   &\quad\quad\quad\left.+
    f_{\bm{k}_1+\bm{q}_2,-\bm{q}_2}^{NN\pi}
    f_{\bm{k}_2+\bm{q}_1-\bm{q}_2,\bm{q}_2}^{N\Delta\pi}
    \right],
\end{align}
and
\begin{align}
    \mathcal{G}_{\bm{k}_2,\bm{q}_1,\bm{k}_3}^{\Delta\pi N}
    &=
    \mathcal{G}_{\bm{q}_1}^{11}
\left[f_{\bm{k}_2,\bm{q}_1}^{\Delta N\pi}
   f_{\bm{k}_3,-\bm{q}_1}^{NN\pi}
   + f_{\bm{k}_2,\bm{q}_1}^{N\Delta\pi}
    f_{\bm{k}_3-\bm{q}_1,\bm{q}_1}^{NN\pi}
   \right]\cr
   &\,\,+
   \mathcal{G}_{\bm{q}_1}^{12}
   \left[
    f_{\bm{k}_2,\bm{q}_1}^{N\Delta\pi}
    f_{\bm{k}_3,-\bm{q}_1}^{NN\pi}
    +
    f_{\bm{k}_2,\bm{q}_1}^{\Delta N\pi}
    f_{\bm{k}_3-\bm{q}_1,\bm{q}_1}^{NN\pi}
   \right],
\end{align}
where $\mathcal{G}_{\bm{q}}^{11}=-(\varepsilon_{\bm{q},b}+U_{bb}n_0)/E_{\bm{q}}^2$ and
$\mathcal{G}_{\bm{q}}^{12}=U_{bb}n_0/E_{\bm{q}}^2$ are the zero-frequency Bogoliubov Green's functions with
$\varepsilon_{\bm{q},b}=q^2/2M_b$ and
$E_{\bm{q}}=\sqrt{\varepsilon_{\bm{q},b}(\varepsilon_{\bm{q},b}+2U_{bb}n_0)}$ .
One can find that the structure of Eq.~\eqref{eq:12} accompanied by two supefluid phonon exchange is similar to the two-pion-exchange three-body force in the momentum space~\cite{epelbaum2008delta}.
In particular, in the limit of $g\ll U_{bc}\sqrt{n_0}$, one can find $U_{\bm{k}_1,\bm{k}_2,\bm{k}_3}(\bm{q}_1,\bm{q}_2)\propto (q_1^2+\xi_{\rm B}^{-2})^{-1}(q_2^2+\xi_{\rm B}^{-2})^{-1}$ where $\xi_{\rm B}=1/\sqrt{4M_bU_{bb}n_0}$ is the healing length of BEC.
While the two-pion-exchange process involves the pion mass $M_\pi$ in the denominator as $(q_1^2+M_\pi^2)^{-1}(q_2^2+M_\pi^2)^{-1}$,
the healing length $\xi_{\rm B}$ plays the role of the inverse of $M_\pi$, which characterizes the range of interaction.
Such a correspondence can also be found in the comparison between the Yukawa one-pion exchange two-body interaction and the two-body interaction mediated by a superfluid phonon in a BEC~\cite{pethick2008bose}.

\begin{figure}[t]
    \centering
    \includegraphics[width=\linewidth]{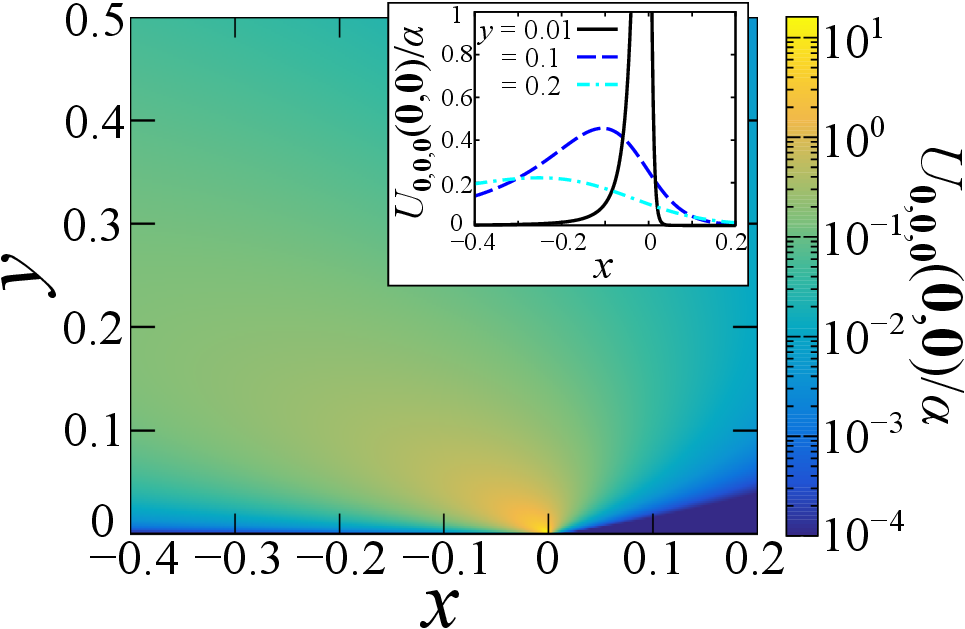}
    \caption{
    Coupling strength $U_{\bm{0},\bm{0},\bm{0}}(\bm{0},\bm{0})$ of the Fujita-Miyazawa-type three-body force in the plane of the dimensionless parameters $x=\{\nu-(U_{bc}+U_{bb})n_0\}/U_{bc}n_0$ and $y=g/U_{bc}\sqrt{n_0}$.
    The inset shows the results at $y=0.01$, $0.1$, and $0.2$.}
     \label{fig:3}
\end{figure}

The magnitude of $U_{\bm{k}_1,\bm{k}_2,\bm{k}_3}(\bm{q}_1,\bm{q}_2)$ is tunable by the external magnetic field via changing $\nu$ in $\mathcal{G}_{\bm{k}_2+\bm{q}_1}^{\Delta}=-1/\xi_{\bm{k}_2+\bm{q}_1,\Delta}$.
To see this, we consider the long-wavelength limit of $U_{\bm{0},\bm{0},\bm{0}}(\bm{0},\bm{0})$, which in turn is relevant for the evaluation of the mean-field energy
$\delta E_3\propto U_{\bm{0},\bm{0},\bm{0}}(\bm{0},\bm{0})n_{N}^3$
in the dilute limit
with the $N$-like polaron number density $n_N$.
Note that the $n_{N}$ dependence of the interaction energy allows us to distinguish $\delta E_3$ from the two-body mean-field energy $\delta E_2\propto n_N^2$.
Indeed, the fermion-mediated two- and three-body interactions were measured by fitting the impurity-density dependence of the speed of sound~\cite{PhysRevLett.131.083003}.
In the dilute limit of the ground-state impurities immersed in the condensate at zero temperature,
we can set $\mu_c=\left[(U_{bc}-U_{bb})n_0+\nu-\sqrt{\{\nu-(U_{bc}+U_{bb})n_0\}^2+4g^2n_0}\right]/2$ such that $\xi_{\bm{k}=\bm{0},N}=0$,
where we used $\mu_b=U_{bb}n_0$.
We thus obtain
\begin{align}
    U_{\bm{0},\bm{0},\bm{0}}(\bm{0},\bm{0}) 
    &=
    \frac{\alpha y^2\left(
    1-{x}/{2}\right)^2}{(x^2+4y^2)^{3/2}}
    \left(\frac{1}{2}-\frac{y^2+x/2}{2\sqrt{x^2+4y^2}}\right)^2,
\end{align}
where $\alpha =
{6U_{bc}^3}/{U_{bb}^2 n_0}$, $x=\{\nu-(U_{bc}+U_{bb})n_0\}/U_{bc}n_0$, and $y=g/U_{bc}\sqrt{n_0}$.

Figure~\ref{fig:3} shows $U_{\bm{0},\bm{0},\bm{0}}(\bm{0},\bm{0}) $ in the plane of $x$ and $y$.
It is found that $U_{\bm{0},\bm{0},\bm{0}}(\bm{0},\bm{0})$ is significantly enhanced near $x\rightarrow 0$ and $y\rightarrow 0$.
Since the interspecies scattering length $a$ can approximately be given by $\frac{2\pi a}{M_{\rm r}}\simeq U_{bc}+\frac{g^2}{\nu}$
for the narrow Feshbach resonance with small $g$~\cite{ohashi2020bcs} and $M_{\rm r}=1/(M_b^{-1}+M_c^{-1})$,
one can see that $U_{\bm{0},\bm{0},\bm{0}}(\bm{0},\bm{0})$
 is enhanced near the resonance ($a\rightarrow \pm\infty$ at $\nu\rightarrow 0$), while there is an additional mean-field shift $(U_{bb}+U_{bc})n_0$.
On the other hand,
$U_{\bm{0},\bm{0},\bm{0}}(\bm{0},\bm{0})$ monotonically decreases when $x$ goes away from the resonant condition.

\begin{figure}[t]
    \centering
    \includegraphics[width=1.0\linewidth]{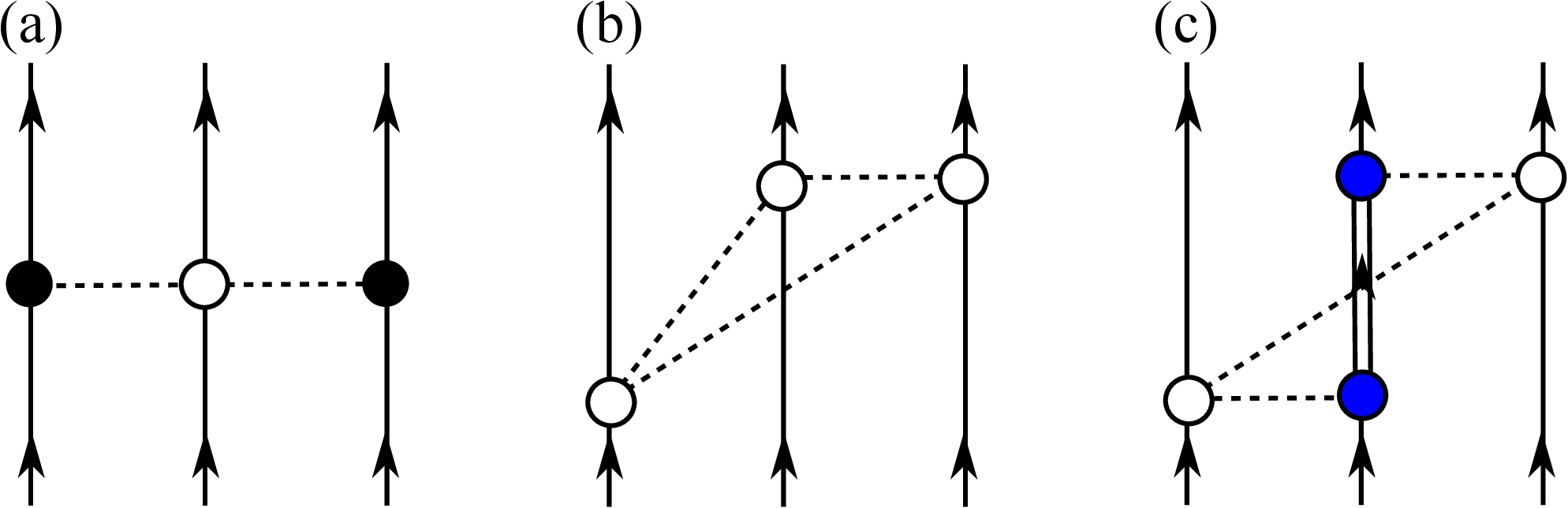}
    \caption{Feynmann diagrams of the other three-body interactions:
    (a) Two-phonon exchange with a two-phonon vertex,
    (b) Three-phonon exchange with three two-phonon vertices,
    (c) Three-phonon exchange with a virtual $\Delta$-like state. 
    The symbols shown in Fig.~\ref{fig:2} are used.}
    \label{fig:4}
\end{figure}

Finally, we discuss
how $\hat{V}_{\rm FM}$ is distinguishable from all the other mediated three-body forces with three and four vertices including two-phonon ones
as shown in Fig.~\ref{fig:4}.
The diagram (a) in the single-channel model was examined theoretically in Ref.~\cite{naidon2020polaron}.
The diagram (b), which has been recently studied in Ref.~\cite{theel2025effective}, is the bosonic counterpart of the three-fermion exchange interaction~\cite{PhysRevA.69.043607,PhysRevA.76.013609,friman2011three,PhysRevA.104.L041302,tajima2021polaron,PhysRevC.104.065801,PhysRevA.109.013319}.
The diagram (c) involves a virtual $\Delta$-like state and a loop of the superfluid phonon propagator.
There are two important differences between $\hat{V}_{\rm FM}$ and these diagrams:
The first one is the $\nu$
dependence of each term.
While the diagrams (a) and (b) are independent of $\nu$, $\hat{V}_{\rm FM}$ and the diagram (c)
depend on $\nu$ through $\xi_{\bm{k},\Delta}$, which enables us to distinguish the two groups by tuning the external magnetic field.
The second difference is the presence or absence 
of two-phonon vertices.
All the diagrams in Fig.~\ref{fig:4} involve at least one two-phonon vertex, which is $O(U_{bc})$ and thus leads to a smaller contribution to the ground-state energy than the one-phonon vertex contribution of $O(U_{bc}\sqrt{n_0})$.
Remarkably, $\hat{V}_{\rm FM}$ is only the term that depends on $\nu$ and does not involve any two-phonon vertex.
This fact is reminiscent of the original Fujita-Miyazawa three-nucleon force that plays an important role in few-body nuclear physics, even compared to the other contributions involving three and more pions~\cite{10.1143/PTP.17.360,fujita1962nuclear}.

{\it Summary.---}
In this work, we have proposed a way to realize the Fujita-Miyazawa-type three-body force among three impurities in an atomic BEC near the interspecies Feshbach resonance.  
In this realization, the coupling strength of the three-body force can be tuned by the external magnetic field, which controls the closed-channel molecular energy.  
Our result is based on a close analogy between dressed impurities
and real nucleons in terms of polaron physics.
The proposed Fujita-Miyazawa-type three-body force can be confirmed experimentally by
attempting the measurements of the equation of state~\cite{desalvo2019observation,PhysRevLett.124.163401,baroni2024mediated,cai2025fermion,PhysRevLett.131.083003}
of the present system as a function of the impurity density
or the theoretical proposals for detecting the interpolaron interactions (e.g., Refs.~\cite{PhysRevLett.112.155301,mistakidis2020many,PhysRevA.102.051302,PhysRevA.103.043334,PhysRevLett.129.233401,PhysRevLett.129.153401}).


{\it Acknowledgments.---}
The authors thank Y. Guo, M. Horikoshi, H. Liang, K. Sekiguchi, Y. Xiao, and H. Yabu for useful discussions.
This work is supported by JSPS KAKENHI for Grants (Nos.\ JP22K13981, JP23K22429, JP23K25864, JP24K06925, JP25K01001) from MEXT, Japan.

\bibliographystyle{apsrev4-2}
\bibliography{reference.bib}

\end{document}